\documentclass{JHEP3}

  \usepackage{latexsym,bm,amsmath,amssymb,amsfonts}
  \usepackage{epsfig,graphics,graphicx}
  \usepackage{slashed}
  \usepackage[latin1]{inputenc}
  \usepackage{mathrsfs}
\usepackage{mathcomp}

  \long\def\comment#1{ }

  \newcommand{\beq}{\begin{eqnarray}}
  \newcommand{\eeq}{\end{eqnarray}}
  
 \def\simge{\mathrel{%
   \rlap{\raise 0.511ex \hbox{$>$}}{\lower 0.511ex \hbox{$\sim$}}}}
\def\simle{\mathrel{
   \rlap{\raise 0.511ex \hbox{$<$}}{\lower 0.511ex \hbox{$\sim$}}}}

\vspace{1.5cm}

\title{\rm \LARGE \bf Heavy quark in an expanding plasma in AdS/CFT}

\author{G. C. Giecold \\Institut de Physique Th\'eorique,
CEA Saclay, CNRS (URA 2306),
 F-91191 Gif-sur-Yvette, France\\
  E-mail: \email{gregory.giecold@cea.fr}}

\abstract{Using the Janik--Peschanski dual to a Bjorken flow, a Langevin equation is derived for a heavy quark in an expanding $\mathcal{N} = 4$ supersymmetric Yang--Mills plasma. Such a plasma is characterized by a proper--time dependence of the temperature and corresponds to a system out of equilibrium. The analysis first focuses on a quark at rest in the plasma comoving frame. The case of a quark moving across a longitudinally expanding plasma is then considered. The two--point functions for the random noise acting on such heavy quark probes are computed.}

\begin{document}

\section{Introduction}

Many problems in physics require going beyond standard Feynman diagrams and S--matrices calculations. In non--equilibrium settings, interactions generally take place in a short time interval and cannot be switched adiabatically as is done e.g. in the LSZ reduction formula for scattering experiments. An asymptotic state might also be out of grasp due to an inherent instability of the system. The initial state is known though, so that $\langle in \mid in \rangle$ matrix elements still provide valuable data. This is at the core of the Schwinger--Keldysh method where the amplitudes are calculated along a path extended in the complex time plane \cite{Keldysh:1964ud, Landsman:1986uw}.
While in non--equilibrium statistical physics the response of a system to a disturbance can often be reduced to real--time Green functions for thermal equilibrium systems, the method of Keldysh Green functions was historically first developed to directly tackle systems out of equilibrium. Equilibrium and non--equilibrium statistical physics are actually formally equivalent when one introduces a contour--ordering to replace the usual time--ordering. See section 2.1.3 of \cite{Landsman:1986uw} and references therein for a more detailed discussion. Non--equilibrium statistical physics is concerned with correlors of the type $\langle \mathcal{O}(t) \rangle = \text{Tr}\left[ \rho \mathcal{O}(t) \right]$ for $t > t_i$, where $\rho$ denotes the distribution for an equilibrium hamiltonian but $\mathcal{O}(t)$ is an operator in a Heisenberg representation with respect to an hamiltonian with an interactions part. Here, $t_i$ refers to an initial time. The standard procedure for obtaining a non--equilibrium state is to consider a state which until $t_i$ was in equilibrium with a reservoir and was thus prepared in some initial conditions. At $t > t_i$ the state is disconnected from the reservoir and interactions are switched on. In fact, unless for fleeting properties of a system out of equilibrium, the dependence on the initial state is rapidly lost due to interactions and the distribution $\rho$ is arbitrary in this case.\\
In \cite{Herzog:2002pc} a prescription for computing Keldysh Green functions in the AdS/CFT correspondence was found and later implemented in \cite{Gubser:2006nz} for computing transverse and longitudinal momentum broadening for a heavy quark, from variations of the underlying Wilson line. Recent works \cite{deBoer:2008gu, Giecold:2009cg, Son:2009vu} explore the Langevin description for a heavy quark from the gauge--string duality. The present paper aims at generalizing this to an expanding plasma. This is a non--equilibrium situation. In particular the medium is characterized by a local temperature whose proper--time dependence obeys a scaling law first devised by Bjorken \cite{Bjorken:1982qr}.
Actually, the recipe of \cite{Herzog:2002pc} for computing real--time correlators in AdS/CFT was later justified in a series of papers by Skenderis and van Rees \cite{Skenderis:2008dh, Skenderis:2008dg, vanRees:2009rw}. See \cite{vanRees:2009rw} for a review and further explanations on how their results reduce to the ingoing boundary condition for bulk fields of \cite{Herzog:2002pc} when the sources are set equal on both boundaries of the Penrose diagrams used in such calculations. 
Besides, their work is amenable to all sorts of initial states and ensembles by switching additional sources in the Euclidean segments of the devised construction.\\
The authors of \cite{Chesler:2008hg} studied horizon formation and thermalization in a non--Abelian plasma resulting from turning on background fields, described by gravitational waves. It would be very interesting to derive transport coefficients such as momentum broadening coefficients for a hard probe from the numerical analysis presented in \cite{Chesler:2008hg} but how this might be achieved is obscured by a lack of hindsight for an evolution out of equilibrium at strong coupling in AdS/CFT. The approach presented in the present paper relies on the leading--order expansion at large times to the Janik--Peschanski dual \cite{Janik:2005zt, Janik:2006gp} to a Bjorken flow and it allows for explicit results.\\
The next section first reviews the work of Kim, Sin and Zahed \cite{Kim:2007ut} and explains how to derive, for a quark at rest in the expanding plasma comoving frame, a Langevin equation. The correlators of the random forces thereof are computed. Section 3 is concerned with a fast quark moving transversally in a strongly coupled $\mathcal{N} = 4$ supersymmetric Yang--Mills plasma experiencing Bjorken flow. The gravity dual corresponds to a string trailing in the Janik--Peschanski background. The dispersions relations, energy loss parameter and momentum broadening coefficients are derived. They exhibit the expected scaling behaviour for the temperature, with no other dependence on the initial thermalization temperature. The method used to compute those quantities in a non--equilibrium, expanding plasma relies on a coordinate change and a particular Fourier--like mode--expansion to map the problem to a situation where the background has a fixed, global temperature.

\section{Transverse and rapidity fluctuations in an expanding plasma and the Langevin description}

In \cite{Janik:2005zt, Janik:2006gp} the gravity dual to a Bjorken flow \cite{Bjorken:1982qr} was derived in a $\tau^{-2/3}$ expansion to the bulk metric in Fefferman-Graham coordinates. The proper--time of an expanding plasma $\tau$ is related, along with the rapidity $y$, to the physical laboratory time $t_{lab}$ and direction of expansion $x^3$ as $t_{lab} = \cosh(y) \tau$, $x^3 = \sinh(y) \tau$. Those parameters are convenient for describing the hydrodynamic regime which takes over after a scenario where typically at proper time $\tau = 0$ two gold nuclei collide at high enough energy that their subsequent evolution leads to a quark--gluon plasma. At proper time $\tau_0$ the resulting plasma is thermalized and its properties are described by Bjorken's hydrodynamic model \cite{Bjorken:1982qr}. The plasma expands along the collision axis. Most useful to the remainder of this paper is the scaling law
\beq\label{Bjorken T}
T^3 \tau^{\alpha} = const,
\eeq 
where $\alpha = 3 v_S^2$. From conformal invariance the sound velocity $v_S$ is set to $1/\sqrt{3}$ and then $\alpha = 1$. In this paper, especially in Section 3, it is assumed that the plasma expands for a sufficiently long period of time that the quark probes a large enough distance $L$ of the quark--gluon plasma. No other possible phase will be considered.\\
 The leading order result in the JP expansion reads
\beq\label{JP metric}
ds^{2} = \frac{R^2}{z_H^2 z^2} \left[-\frac{(1-w^4)^2}{(1+w^4)} d\tau^2 + (1+w^4) \left[ \tau^2 dy^2 + dx_{\perp}^2 \right] + z_H^2 dz^2 \right],
\eeq
with $w = \frac{z}{(\tau / \tau_0)^{1/3}} \epsilon^{1/4}$, $\epsilon = (\pi T_0)^4 / 4$ and where $T_0 = \frac{1}{\pi z_H}$ is the Hawking temperature.\\
The picture that emerges is that of a black hole whose horizon is moving away from the boundary. The coordinate change, $\frac{t}{t_0} = \frac{3}{2} (\frac{\tau}{\tau_0})^{2/3}$, $u(t,z) = \frac{2 w^2}{1+w^4}$ -- after discarding the non--diagonal components which ensue, as they are subleading in the $\tau$ expansion -- yields \cite{Kim:2007ut} 
\beq\label{Metric}
ds^{2} = \frac{R^{2}}{z_H^{2} u} \left( -f dt^{2} + \frac{4t^2}{9} dy^2 + \frac{3t_0}{2t} dx_{\perp}^{2} + z_H^2 \frac{du^{2}}{4 u f} \right), \eeq
where $f(u) = 1 -  u^{2}$.\\
 The above form of the metric proves convenient as it converts a time--dependent problem into a setting where the usual recipe for extracting dual gauge theory correlators from fields in a AdS--Schwarzschild black hole background applies.\\
The time--dependent transverse and rapidity components of the metric are accounted for by Fourier--Hankel transformations. The study of transverse string fluctuations was carried out in \cite{Kim:2007ut} where the corresponding momentum broadening coefficient and diffusion coefficients were found for a heavy quark probe at rest in the plasma co--moving frame.\\
 The remainder of this section generalizes this to the rapidity fluctuations as well. Moreover, the Kubo--Martin--Schwinger formula relating the retarded and symmetric correlators is derived. The construction of Schwinger--Keldysh propagators in AdS/CFT first devised in \cite{Herzog:2002pc} and later justified in \cite{Skenderis:2008dh, Skenderis:2008dg, vanRees:2009rw} thus holds. This then ensures for the existence of a Langevin description -- which was merely postulated in \cite{Kim:2007ut}.\\
 The Nambu--Goto action
\beq\label{NG}
S_{NG} = - \frac{1}{2\pi \alpha'} \int d^2\sigma \sqrt{-g}, \ \ \ \ \ g_{\alpha \beta} = G_{\mu \nu} \partial_{\alpha}X^{\mu} \partial_{\beta}X^{\nu}, 
\eeq 
can be expanded to quadratic order in the transverse and rapidity fluctuations $\delta X^{1,2} = \xi^{1,2}(t,u)$ and $\delta y(t,u)$ in the background specified by the target--space metric components $G_{\mu \nu}$ of (\ref{Metric}). This provides
\begin{align}\label{NG quadra}
S_{NG} =& - \frac{\sqrt{\lambda} T_0}{4} \int dt \ du \frac{1}{u^{3/2}} + \frac{\sqrt{\lambda} T_0}{8} \int dt \ du \left( \frac{3t_0}{2t} \right) \sum_{i = 1,2} \left[ \frac{(\partial_t \xi^{i})^2}{u^{3/2} f(u)} - (2 \pi T_0)^2 \frac{f(u)}{u^{1/2}} (\partial_u \xi^{i})^2 \right] \nonumber \\ &+ \frac{\sqrt{\lambda} T_0}{8} \int dt \ du \left( \frac{4t^2}{9} \right) \left[ \frac{(\partial_t \delta y)^2}{u^{3/2} f(u)} - (2 \pi T_0)^2 \frac{f(u)}{u^{1/2}} (\partial_u \delta y)^2 \right].
\end{align}
Here $\lambda = R^2/\alpha' >> 1$, so that string loop corrections are negligible at this order and computations at the two--derivatives supergravity level are reliable. The action (\ref{NG quadra}) is the same as in a static black hole background apart from overall time--dependent factors. The equations of motion are
\beq\label{EOM}
\left[ \partial_t^2 - \frac{1}{t} \partial_t + 2 \pi^2 T_0^2 f(u) (1 + 3 u^2) \partial_u - (2 \pi T_0)^2 u f(u)^2 \partial_u^2 \right] \xi^{i} = 0, \ \ \ i = 1,2\\
\eeq
along with
\beq
\left[ \partial_t^2 + \frac{2}{t} \partial_t + 2 \pi^2 T_0^2 f(u) (1 + 3 u^2) \partial_u - (2 \pi T_0)^2 u f(u)^2 \partial_u^2 \right] \delta y = 0.
\eeq
Expanding in a basis defined by Hankel functions
\beq\label{Hank exp}
\left\{ 
\begin{array}{rl}
\xi^{i}(t,u) = \int_{-\infty}^{\infty} \frac{d\omega}{2\pi} \sqrt{\frac{i\pi \omega}{2}} t H_1^{(2)}(\omega t) \Psi_{\omega}(u) \tilde{\xi_0}(\omega) ;\\
\delta y(t,u) = \int_{\infty}^{\infty} \sqrt{\frac{i\pi \omega}{2}} (\frac{-i}{\sqrt{t}}) H_{1/2}^{(2)}(\omega t) \Phi_{\omega}(u) \tilde{\delta y_0}(\omega),
\end{array} \right.
\eeq
yields 
\beq\label{EOM Psi Phi}
\left[ \partial_u^2 - \frac{3 u^2 + 1}{2 u f(u)} \partial_u + \frac{\mathfrak{w}^2}{4 u f(u)^2} \right] \left(
\begin{array}{ccc}
\Psi_{\omega} \\
\Phi_{\omega}
\end{array} 
\right) (u) = 0, \ \ \ \mathfrak{w} = \frac{\omega}{\pi T_0}, \eeq
where $\Psi_{\omega}$ and $\Phi_{\omega}$ are normalized to unity at $u = 0$.
Inserting (\ref{Hank exp}) into (\ref{NG quadra}), using the equations of motion and integrating by parts gives
\beq\label{S bndry}
S_{bndry} = \frac{3 \pi^2 \sqrt{\lambda} T_0^3 t_0}{4} \sum_{i = 1,2} \int dt \ \frac{f(u)}{\sqrt{u} t} \xi^{i} \partial_u \xi^{i} (t,u) \mid_{u = 0}^{u = 1} \nonumber \\
+ \frac{2 \pi^2 \sqrt{\lambda} T_0^3}{9} \int dt \ t^2 \frac{f(u)}{\sqrt{u}} \delta y \partial_u \delta y (t,u) \mid_{u = 0}^{u = 1},
\eeq
One then appeals to the approximate completeness relation
\beq\label{completeness}
- \frac{1}{4} \int_{-\infty}^{\infty} dt \ t H_{\nu}^{(2)}(\omega t) H_{\nu}^{(2)}(-\omega' t) \simeq \frac{1}{\omega} \delta(\omega - \omega'),
\eeq
which stems from the exact relation $\int_0^{\infty} dt \ t J_{\nu}(\omega t) J_{\nu}(\omega' t) = \frac{1}{\omega} \delta(\omega - \omega')$ for Bessel functions.
One can argue that (\ref{completeness}) is a fair approximation given that the dominant contributions in the integrals come from the late time region and that the Janik--Peschanski metric is defined as a large $\tau$ inverse expansion.\\
As a result, the following expressions for the retarded Green functions hold
\beq\label{G R}
\left\{
\begin{array}{ll}
G_{R,\perp} (\omega) = \left[- \frac{3 \pi^2 \sqrt{\lambda} T_0^3 t_0}{2} \right] \left[\frac{f(u)}{\sqrt{u}} \Psi_{-\omega}(u) \partial_u \Psi_{\omega}(u) \right]_{u = 0} ;\\
G_{R, \delta y} (\omega) = \left[ \frac{4 \pi^2 \sqrt{\lambda}T_0^3}{9} \right] \left[\frac{f(u)}{\sqrt{u}} \Phi_{-\omega}(u) \partial_u \Phi_{\omega}(u) \right]_{u = 0}.
\end{array} \right.
\eeq
 In the following, it is checked explicitly that the symmetrized Wightman functions $G_{sym}(\omega)$ are related to the corresponding retarded correlators by a Kubo--Martin--Schwinger (KMS) relation \cite{LeBellac:1996CUP, Landsman:1986uw}
\beq\label{KMS}
G_{sym}(\omega) = - \coth(\frac{\omega}{2 T_0}) Im \ G_R(\omega).
\eeq
It involves the temperature $T_0$, which is the initial, thermalization temperature in the original Bjorken frame. The following illustrates how the proper--time dependent temperature appears in the 2--point functions for this frame, from the Green functions computed in the $\{t-u\}$ system.\\ 
For this purpose let us follow the usual prescription as it appears in \cite{CasalderreySolana:2006rq, CasalderreySolana:2007qw, Giecold:2009cg, Gubser:2006nz, Herzog:2002pc, Son:2009vu} and expand a general solution in the right and left quadrants of the black hole background (\ref{Metric}), whose Kruskal diagram is the same as for a AdS--Schwarzschild black hole : 
\beq\label{R L exp}
\left\{
\begin{array}{ll}
\Upsilon_{R,\ \omega}(u) = A(\omega) \Psi^H_{\omega,\ in}(u) + B(\omega) \Psi^H_{\omega,\ out}(u) ;\\
\Upsilon_{L,\ \omega}(u) = C(\omega) \Psi^H_{\omega,\ in}(u) + D(\omega) \Psi^H_{\omega,\ out}(u),
\end{array} \right.
\eeq
$\Upsilon_{\omega}(u)$ denoting collectively $\Psi_{\omega}(u)$ or $\Phi_{\omega}(u)$ from (\ref{EOM Psi Phi}).\\
$\Psi^H_{\omega,\ in}(u)$ and $\Psi^H_{\omega, \ out}(u)$ form a basis of two independent wave--functions whose expansion near the horizon at $u = 1$ is, up to $\mathcal{O}(\omega^2)$ terms
\beq\label{Psi basis}
\left\{
\begin{array}{ll}
\Psi^H_{\omega,\ in} = (1 - u^2)^{-i \frac{\mathfrak{w}}{4}} \left[1 + \frac{i \mathfrak{w}}{8} (\pi - 4 \tan^{-1}(\sqrt{u}) - 6 \log(2) + 2 \log(1+u)(1+\sqrt{u})^2) \right] \ ; \\
\Psi^H_{\omega,\ out} = \Psi^{H*}_{\omega,\ in}.
\end{array} \right.
\eeq
The Kruskal coordinates are cast in the form $U = - \frac{1}{2 \pi T_0} e^{-2 \pi T_0 (t - r_*)}$, $V = \frac{1}{2 \pi T_0} e^{2 \pi T_0 (t + r_*)}$.\\
$r_* = \frac{1}{4 \pi T_0} \left[ 1 + \log(\frac{1}{u} - 1) \right]$ denotes the tortoise coordinate.\\
From
\beq\label{U V H exp}
\left\{
\begin{array}{lll}
(-U)^{\frac{i\mathfrak{w}}{2 \pi T_0}} \simeq (1 - u)^{i\mathfrak{w}/4} e^{-i \omega t} (\frac{1}{2 \pi T_0})^{i \mathfrak{w}/2} e^{-i \mathfrak{w}/4} \left[ 1 + (1-u) \frac{i \mathfrak{w}}{2} \right] \ ;\\ 
(V)^{-\frac{i\mathfrak{w}}{2 \pi T_0}} \simeq (1 - u)^{-i\mathfrak{w}/4} e^{-i \omega t} (\frac{1}{2 \pi T_0})^{- i \mathfrak{w}/2} e^{-i \mathfrak{w}/4} \left[ 1 - (1-u) \frac{i \mathfrak{w}}{2} \right] \ ;\\
H_{\nu}^{(2)}(\omega t) \simeq \sqrt{\frac{2}{\pi \omega t}} e^{-i(\omega t - \pi \nu / 2 - \pi / 4)},\ \ \  \mid \omega t \mid \rightarrow \infty,
\end{array} \right.
\eeq
near the horizon, one obtains the following behaviour for the modes satisfying the equations of motion :
\beq\label{Psi in out expansion}
\left\{
\begin{array}{ll}
\sqrt{\frac{i \pi \omega}{2}} t H_1^{(2)} (\omega t) \Psi^H_{\omega,\ in } (u \simeq 1) \simeq \sqrt{\log(\frac{-V}{U})} e^{-\frac{i\mathfrak{w}}{2} \log(V)} \ ; \\
\sqrt{\frac{i \pi \omega}{2}} t H_1^{(2)} (\omega t) \Psi^H_{\omega,\ out } (u \simeq 1) \simeq \sqrt{\log(\frac{-V}{U})} e^{-\frac{i\mathfrak{w}}{2} \log(-U)},
\end{array} \right.
\eeq
and
\beq\label{Phi in out expansion}
\left\{
\begin{array}{ll}
\sqrt{\frac{i \pi \omega}{2}} (\frac{-i}{\sqrt{t}}) H_1^{(2)} (\omega t) \Psi^H_{\omega,\ in } (u \simeq 1) \simeq \frac{1}{\log(\frac{-V}{U})} e^{-\frac{i\mathfrak{w}}{2} \log(V)} \ ; \\
\sqrt{\frac{i \pi \omega}{2}} (\frac{-i}{\sqrt{t}}) H_1^{(2)} (\omega t) \Psi^H_{\omega,\ out } (u \simeq 1) \simeq \frac{1}{\log(\frac{-V}{U})} e^{-\frac{i\mathfrak{w}}{2} \log(-U)},
\end{array} \right.
\eeq
 The conditions at the horizon used in \cite{Herzog:2002pc} amount to the analyticity of the infalling modes in the lower $V$ complex plane (which guarantees that such modes carry positive energy). Similarly, they guarantee that the outgoing solutions are of negative energy, hence analytic in the upper $U$ plane. A full justification of this recipe and a generalization to a broader framework for computing real--time correlators in the gauge/gravity correspondence appears in \cite{Skenderis:2008dh, Skenderis:2008dg, vanRees:2009rw}.
One can generalize and extend the transformation from the right quadrant ($U < 0$, $V>0$) to the left quadrant ($U > 0$, $V<0$) to $V \rightarrow \mid V \mid e^{-i \theta}$, $-U \rightarrow \mid U \mid e^{-i (2 \pi - \theta)}$ \cite{Son:2009vu}, where $\theta$ was naturally set to $\pi$ in \cite{Herzog:2002pc}. In the case at hands, $\theta = 0 \ mod \ [\pi]$ is most convenient. $\theta = 0$ leads to a treatment in terms of retarded and advanced wave--functions $\Upsilon_a = \Upsilon_R - \Upsilon_L$, $\Upsilon_r = \frac{\Upsilon_R + \Upsilon_L}{2}$. The current problem is thus amenable to the same discussion as in \cite{Giecold:2009cg, Son:2009vu}. Following the analysis expounded in those references, 
\beq\label{}
\left(
\begin{array}{ccc}
C \\
D
\end{array} \right)(\omega) = \left( \begin{array}{ccc} 1 & 0 \\
0 & e^{\omega / T_0} \\ \end{array} \right) \left( \begin{array}{ccc}
A\\
B\end{array} \right)(\omega), 
\eeq
and from here on the recipe for obtaining a Langevin equation applies :
\begin{align}\label{Sbndry Lang}
i S_{bndry, \perp} =& - i \int \frac{d\omega}{2\pi} x^h_{a,\perp}(-\omega) \left[ G^h_{R,\perp}(\omega) \right] x^h_{r,\perp}(\omega, \perp) \nonumber \\
&- \frac{1}{2} \int \frac{d\omega}{2\pi} x^h_{a,\perp}(-\omega) \left[ G_{sym,\perp}(\omega) \right] x^h_{a,\perp}(\omega),
\end{align}
and similarly for the rapidity sector with -- e.g. for transverse fluctuations --
\begin{align}\label{bndry G R}
G^h_{R, \perp} (\omega) &= - \frac{3\pi \sqrt{\lambda} T_0^2 t_0}{4} i \omega, \nonumber \\
&= - i \omega \eta_{\perp}, 
\end{align}
and
\begin{align}\label{bndry G sym}
G^h_{sym, \perp} (\omega) &= \frac{3\pi^2\sqrt{\lambda}T_0^3 i t_0}{2} \frac{(1+2n(\omega))}{2} \left[ \frac{f(u)}{\sqrt{u}}\partial_u(\Psi^H_{\omega, \ in} - \Psi^H_{\omega, \ out}) \right]_{u = 1}, \nonumber \\ &= -(1+2n(\omega))Im \ G_{R, \perp} (\omega).
\end{align}
$n(\omega)$ denotes the thermal distribution at temperature $T_0$. In the original Bjorken variable, this is the temperature at the thermalization time $\tau_0$. In the Bjorken frame, the temperature subsequently decreases according to the scaling law (\ref{Bjorken T}). Yet, the above analysis was performed at a single temperature $T_0$. How should one possibly expect to gain knowledge of 2--point functions in an expanding plasma ? The change of coordinates that we made allows for an analysis where the plasma local temperature is kept at $T_0$. The time dependence is indeed transferred only to the transverse and velocity coordinates components of the metric, while the time and radial components take on the same form as for an AdS--Schwarzschild black hole with temperature $T_0$. In the following, we show how the physical temperature $T(\tau)$ at proper--time $\tau$ makes its way in the coefficients of the Langevin description.\\
The bulk picture of Brownian motion \cite{Giecold:2009cg, Son:2009vu} leads to a stochastic equation with random noise $\xi$ for the horizon endpoint of the string :
\beq\label{stochstic ndpoint}
\left\{
\begin{array}{ll}
T_{\perp}(u_h) \partial_u x_{r,\perp}(\omega, u) + \xi^h_{\perp}(\omega) = - i \omega \eta_{\perp} x^h_{r,\perp}(\omega), \\ 
\langle \xi^h_{\perp}(-\omega) \xi^h_{\perp}(\omega) \rangle = \eta_{\perp} \omega [1+2n(\omega)].
\end{array} \right.
\eeq
$T_{\perp}(u) = \frac{3\pi^2\sqrt{\lambda}T_0^3t_0}{2}\frac{1-u^2}{\sqrt{u}}$ is the local tension in the string. In the long time limit where the relevant scales are large with respect to the heavy quark relaxation time, this term is negligible as the string appears straight and the bulk has no effect on the stretched horizon. Therefore the equation of motion for the horizon endpoint is
\beq\label{EOM horizon}
\frac{dx^h_{\perp}}{dt} \simeq \frac{\xi^h}{\eta_{\perp}}
\eeq
and similarly for the boundary endpoint. Going from (\ref{stochstic ndpoint}) to (\ref{EOM horizon}) requires the completeness relation. Also, terms of order $\mathcal{O}(1/\sqrt{t})$ were discarded in $d(\eta_{\perp}\sqrt{t} x^h_{\perp})/dt$.\\
Besides, using the inverse of (\ref{completeness}),
\begin{align}\label{xi xi correl}
\langle \xi^h_{\perp}(t_1) \xi^h_{\perp}(t_2) \rangle &= -\frac{1}{4} \int_{-\infty}^{\infty} d\omega \omega H_1^{(2)}(\omega t_1) H_1^{(2)}(- \omega t_2) \langle \xi^h_{\perp}(\omega) \xi^h_{\perp}(-\omega) \rangle\nonumber \\
&\simeq \frac{3 \pi \sqrt{\lambda} T_0^3 t_0}{2} \frac{1}{(t_1 + t_2)/2} \delta(t_1 - t_2), \nonumber \\ &= K_{\perp}(t_1, t_2).
\end{align}
Use has been made of $t_{1,2} >> 1$, as appropriate from the JP asymptotic condition. Besides, $\sqrt{t_1 t_2} = \sqrt{\mathcal{T}-\frac{s^2}{4}}$, where $\mathcal{T} = \frac{t_1 + t_2}{2}$, $s = t_1 - t_2$, and the conditions $\mathcal{T} >> 1$, $s << 1$ were then invoked.\\
In a Langevin description which gives the dynamics of the heavy quark propagating in the expanding plasma  
\beq\label{Lang t var}
\begin{array}{lll}
\frac{dp_i}{dt} = F_i^L + F_i^T,\\
\langle F_i^L (t_1) F_j^L(t_2) \rangle = \hat{p}_i \hat{p}_j K_L(t_1,t_2),\\
\langle F_i^T (t_1) F_j^T(t_2) \rangle = (\delta_{ij} - \hat{p}_i \hat{p}_j) K_T(t_1,t_2).
\end{array}
\eeq
Switching to the proper--time coordinate, each force component comes with an additional factor of $\sqrt{3t_0 / 2 t}$ : $ F_i^{L,T}(\tau_{1,2}) = \sqrt{\frac{3 t_0}{2 t_{1,2}}} F_i^{L,T}(t(\tau_{1,2}))$.\\
Taking care of the Dirac distribution transformation law under coordinate change, this yields, e.g. for the transverse force,
\begin{align}\label{XXXXX}
\langle F_i^T(\tau_1) F_j^T(\tau_2) \rangle &= (\frac{3t_0}{2t_1})^2 \pi \sqrt{\lambda} T_0^3 (\frac{\tau_1}{\tau_0})^{1/3} \delta(\tau_1 - \tau_2) (\delta_{ij} - \hat{p}_i \hat{p}_j) \nonumber \\
&= \pi \sqrt{\lambda} T^3(\tau_1) \delta(\tau_1 - \tau_2) (\delta_{ij} - \hat{p}_i \hat{p}_j). 
\end{align}
In the Bjorken frame the force correlator thus exhibits a simple scaling law on the temperature with no explicit dependence on the initial temperature $T_0$. The initial condition on the temperature is then partially washed out. As a landmark of adiabatic evolution, though, it is still hidden in the scaling law for the local temperature.

\section{Trailing string in the BF background} 

We now turn to the case of a heavy quark probe moving transversally at some average velocity through an expanding strongly--coupled $\mathcal{N} = 4$ SYM plasma. Suppose that after the hydrodynamic regime has settled, a heavy quark is created among the debris of the collision and starts propagating through the thermalized state of matter with vanishing longitudinal momentum, which means it lies at rapidity $y = 0$. Hence, the proper time parameter $\tau$ in the comoving frame measures the physical time elapsed since the probe departed. The quark will hit subsequent layers of matter at different cooling temperatures and densities. In particular, the temperature is described by the scaling law (\ref{Bjorken T}).\\
 In the context of weakly--coupled quantum chromodynamics, the authors of \cite{Baier:1998yf} studied the energy loss and momentum broadening for such a probe created either inside or coming from outside of such an expanding plasma. Their analysis relied on perturbation theory. From $\hat{q}(\tau) = \rho(\tau) \int d^2\vec{q}_{\perp} \vec{q}_{\perp}^2 \frac{d\sigma}{d^2\vec{q}_{\perp}}$, with $\rho(\tau)$ the position--dependent density of the medium, which entails $\hat{q}(\tau) = \hat{q}(\tau_0)(\frac{\tau_0}{\tau})^{\alpha}$, they found an increase in the rate of energy loss compared to their results in a static medium \cite{Baier:1998yf} :
\beq\label{BDMPS loss}
- \frac{dE}{dx_{\perp}} = \frac{2}{2-\alpha} \left(-\frac{dE}{dx_{\perp}} \right)_{\mid_{static}},
\eeq
in case the quark is produced inside the medium. It should also be noted that theirs is a finite--extent plasma, unlike the one described by the JP dual that is investigated below.\\
As pointed out in \cite{Baier:2002tc}, an expanding medium amounts to an effective transport coefficient $\hat{q}_{eff}(L)$ which would be equivalent to a jet--quenching coefficient in a static plasma :
\begin{align}\label{q hat Baier}
\hat{q}_{eff}(L) &= \frac{2}{L^2} \int_{\tau_0}^{L} d\tau \ (\tau - \tau_0) \ \hat{q}(\tau) \nonumber \\
                        &\simeq \frac{2}{2 - \alpha} \hat{q}(L), 
\end{align}
as the limit $\tau_0 \rightarrow 0$ is taken in much of these studies. The coefficient $\hat{q}(L)$ is evaluated at the temperature $T(L)$ probed by the quark after it has travelled a distance $L$ through the cooling medium.\\
 We would like to learn what happens at strong coupling, despite the difference in the mechanism for energy loss from the one that prevails at weak coupling, in a static plasma, as emphasized, e.g., in \cite{Dominguez:2008vd}. The discussion focuses on the rate of energy loss and, in the final part, on the momentum broadening coefficients.\\
 The starting point is the gravitational dual to the Bjorken flow, the JP metric \cite{Janik:2005zt, Janik:2006gp}. One would like to check if a similar enhancement exists and, besides if independence of the transport and momentum broadening coefficients on the thermalization temperature $T_0$, which was indeed qualified as `remarkable' by \cite{Baier:1998yf}, is observed.\\
It would seem appropriate to start with the ansatz $X^1(\tau,z) = v\tau + \zeta(z)$ $(*)$ for the trajectory of the string and its quark boundary endpoint, so as to gather information on the drag force and momentum broadening coefficients experienced by a heavy quark moving a velocity $v$ in the plasma proper frame at strong coupling. For convenience, we defined $z$ as $\sqrt{u}$. It should not be mistaken with the $z$--variable from the starting JP metric. In the following, the background is provided by the tamed form of the metric (\ref{Metric}). The initial JP metric is far less pliable to tractable computations.\\ 
 However consider instead a different ansatz 
\beq\label{X trail}
X^1 (t, z) = v t + \zeta (z),
\eeq
and momentarily defer a discussion on the difficulties one would have run into, had one chosen to work with the proper--time parameter directly. 
The ansatz (\ref{X trail}) yields
\beq\label{g rho}
\sqrt{-g} = \frac{R^2}{z_H^2 z^2} \sqrt{-\tilde{g}}, \ \ \ \sqrt{-\tilde{g}} = \sqrt{z_H^2(1 - \frac{3t_0}{2t} \frac{v^2}{f(z)}) + \frac{3t_0}{2t} f(z) (\partial_z\zeta)^2}, \eeq
and the equation of motion
\beq\label{EOM trail}
\frac{3t_0}{2t} \frac{f}{z^2} \partial_z \zeta = C \sqrt{-\tilde{g}}.
\eeq
Inserting this implicit expression for the derivative of $\zeta$ and solving for $\sqrt{-g}$ gives
\beq\label{g tilde explicit}
(\sqrt{-\tilde{g}})^2 = z_H^2 \frac{1 - \frac{3t_0}{2t} v^2 - z^4}{1- \left[ 1 + C^2/\frac{3t_0}{2t} \right] z^4}.
\eeq
In order to ensure that $(-g)$ stays positive everywhere on a string that extends from the horizon to the boundary, both numerator and denominator must change sign at the same point (note that $t$ starts at $t_0$). Hence 
\beq\label{C expr}
C = \pm \frac{\frac{3t_0}{2t}v}{\sqrt{1 - \frac{3t_0}{2t}v^2}},
\eeq
and $\partial_z \zeta = \pm v z_H \frac{z^2}{f}$, which is integrated to
\beq\label{X trail 2}
X^1 (t,z) = X^1_0(t,z) = x^1_0 + v t \mp \frac{v z_H}{2} \left[ \tan^{-1}(z) + \log\sqrt{\frac{1-z}{1+z}} \right].
\eeq
In the subsequent discussion the $+$ sign in (\ref{X trail 2}) is always assumed.
 Starting with $(*)$ would result in $X^1(\tau,z) = x_0^1 + v\tau \mp \frac{v z_H}{2} (\frac{\tau}{\tau_0})^{1/3} \left[ \tan^{-1}(z) + \log\sqrt{\frac{1-z}{1+z}} \right]$.
An extra proper time dependence is forced on $\zeta(z)$ from the value taken by $C$ in the process. This is in contradiction with $(*)$. Note that having to work with (\ref{X trail 2}) instead of a linear motion in the plasma proper frame raises no problem if one accepts to momentarily set aside a picture of the quark in Bjorken variables ; actually the following discussion establishes how a trajectory for a quark moving with constant velocity with respect to the proper--time variable appears.

\subsection{Dispersion relations and drag force}

This section investigates the way the dispersion relations and the drag acting on the quark are modified by the changing properties of the plasma. A similar analysis was performed for a string trawling an AdS--Schwarzschild black hole in \cite{Herzog:2006gh}.\\
It is shown that the dispersions relations take on their usual expressions only after a change of reference frame to the starting Bjorken variables is performed. This then leads to the identification of a term responsible for energy loss.\\
The general expressions for the canonical momentum densities to an open string in a background specified by $G_{\mu \nu}$ are
\beq\label{Pi density 0}
\pi^0_{\mu} = - \frac{1}{2 \pi \alpha'} G_{\mu \nu} \frac{(\dot{X}.X^{'})(X^{\nu})^{'} - (X^{'})^2 (\dot{X}^{\nu})}{\sqrt{-g}},
\eeq
\beq\label{Pi density 1}
\pi^1_{\mu} = - \frac{1}{2 \pi \alpha'} G_{\mu \nu} \frac{(\dot{X}.X^{'})(\dot{X}^{\nu}) - (\dot{X})^2 (X^{\nu})^{'}}{\sqrt{-g}}.
\eeq
For a string trailing in a JP background massaged to the metric (\ref{Metric}) this reduces to
\beq\label{Pi 0 trail}
\left\{
\begin{array}{ll}
\pi^0_{t} = - \frac{\sqrt{\lambda} T_0}{2} \frac{1}{\sqrt{1-\frac{3t_0}{2t}v^2}} \frac{ \left[ 1 - (1-\frac{3t_0}{2t}v^2)z^4 \right] }{z^2 f(z)},\\
\pi^0_{x^1} = \frac{\sqrt{\lambda} T_0}{2} \frac{\frac{3t_0}{2t} v}{\sqrt{1-\frac{3t_0}{2t}v^2}} \frac{1}{z^2 f(z)},
\end{array} \right.
\eeq
\beq\label{Pi 1 trail}
\left\{
\begin{array}{ll}
\pi^1_{t} = \frac{\pi \sqrt{\lambda} T_0^2}{2} \frac{\frac{3t_0}{2t} v^2}{\sqrt{1-\frac{3t_0}{2t}v^2}} = \frac{\pi \sqrt{\lambda} T(\tau)^2}{2} \frac{v^2}{\sqrt{1 - \frac{3t_0}{2t}v^2}},\\
\pi^1_{x^1} = \frac{\pi \sqrt{\lambda} T(\tau)^2}{2} \frac{v}{\sqrt{1 - \frac{3t_0}{2t}v^2}}.
\end{array} \right.
\eeq
Integrating along the string, the resulting total energy and momentum, $E = - \int d\sigma \pi^0_t$, $p = \int d\sigma \pi^0_{x^1}$, read
\beq\label{Tot E trail}
E = \frac{\sqrt{\lambda}T_0}{2} \frac{1}{\sqrt{1-\frac{3t_0}{2t}v^2}} \left[ z_m^{-1} - z_h^{-1} + \frac{3t_0}{2t}v^2 \Lambda(z_h) \right],
\eeq
\beq\label{Tot p trail}
p = \frac{\sqrt{\lambda}T_0}{2} \frac{\frac{3t_0}{2t}v}{\sqrt{1-\frac{3t_0}{2t}v^2}} \left[ z_m^{-1} - z_h^{-1} + \Lambda(z_h) \right],
\eeq
where 
\beq\label{Lambda z m}
\Lambda(z_h) = \frac{1}{4} \left[ 2 \ \tan^{-1}(z_m) - 2 \ \tan^{-1}(z_h) + \log\frac{(1-z_m)(1+z_h)}{(1+z_m)(1-z_h)} \right].
\eeq
This compares with eq.(3.21) in \cite{Herzog:2006gh}.\\
The total energy and momentum diverge due to their contributions close to the horizon, i.e. as the cut--off $z_h \rightarrow 1$.\\
The total energy exhibits a contribution $\gamma(t) E_{straight} = 1/\sqrt{1-\frac{3t_0}{2t}v^2} E_{straight}$ identified with the boosted static energy to a frame moving at velocity $v$, where $E_{straight} = \frac{R^2}{2\pi \alpha' z_H}(z_m^{-1}-z_h^{-1})$.\\
Hence the dispersion relation
\beq\label{Dispersion relation}
\left\{
\begin{array}{ll}
E = \gamma(t) \frac{\sqrt{\lambda}T_0}{2} \left[ z_m^{-1} - z_h^{-1} \right] + \frac{1}{v} \frac{dE}{dt} \Delta x^1(z_h), \\
p = \gamma(t) \frac{\sqrt{\lambda}T(\tau)}{2} \sqrt{\frac{3t_0}{2t}} v \left[ z_m^{-1} - z_h^{-1} \right] + \frac{1}{v} \frac{dp}{dt} \Delta x^1(z_h). 
\end{array} \right.
\eeq
$\Delta x^1(z_h)$ is defined as
\beq\label{Delta x}
\Lambda(z_h) = \frac{1}{z_H} \mid \frac{\Delta x^1(z_h)}{v} \mid,
\eeq
with $dE/dt = \pi^1_t$, $dp/dt = -\pi^1_x$.\\
 The square root appearing in the expression for $p$ could potentially spoil the interpretation of these formulas as providing the energy and momentum for a quark moving at velocity $v$ through the plasma. 
Note however that going to the co--moving frame, (\ref{Dispersion relation}) reads
\beq\label{Disp rel}
\left\{
\begin{array}{ll}
\tilde{E} \simeq \gamma(\tilde{v}) \frac{\sqrt{\lambda} T(\tau)}{2} \left[ z_m^{-1} - z_h^{-1} \right] + \frac{1}{\tilde{v}} \frac{d\tilde{E}}{d\tau} \Delta x^1(z_h), \\
\tilde{p} = \gamma(\tilde{v}) \frac{\sqrt{\lambda} T(\tau)}{2} \tilde{v} \left[ z_m^{-1} - z_h^{-1} \right] + \frac{1}{\tilde{v}} \frac{d\tilde{p}}{d\tau} \Delta x^1(z_h), 
\end{array} \right.
\eeq
where $\tilde{v} = \frac{\partial X^1}{\partial \tau} = \sqrt{\frac{3t_0}{2t}}v$ is the speed of a particle moving with constant velocity in the plasma co--moving frame, with the same trajectory as the heavy quark probe described according to (\ref{X trail 2}). Terms of order $\mathcal{O}(\tau^{-4/3})$ have been discarded. This is legitimate given the JP asymptotic condition and the background metric coefficients being actually leading order contributions to an expansion in $\tau^{-2/3}$.\\
It is now straightforward to derive the drag coefficient :
\beq\label{dp dtau}
\frac{d\tilde{p}}{d\tau} = - \eta \tilde{p}, \ \ \ \eta = \frac{\pi \sqrt{\lambda} T^2(\tau)}{2M},
\eeq
which displays the same form as in \cite{Herzog:2006gh}, with the proper--time dependence of the temperature in an expanding plasma now taken into account.\\
This marks a difference in the energy loss mechanism in QCD from the one in a $\mathcal{N}=4$ SYM plasma at strong coupling. In perturbative QCD the energy loss is dominated by induced radiation of gluons. The transverse momentum of those gluons is high enough that the coupling $\alpha_s$ at this scale is weak, allowing for a perturbative calculation for a parton energy loss :
\beq\label{Delta E q hat}
\Delta E = \frac{1}{4}\alpha_s C_R \hat{q} \frac{L^{-2}}{2}.
\eeq 
$L^{-}$ stands for the path length of the parton in the plasma. $\hat{q}$ keeps track of the nonperturbative soft interactions between emitted gluons and the medium and between the emitting parton and the plasma.\\
While in QCD the average loss of energy (\ref{Delta E q hat}) from a fast parton has at most a logarithmic dependence on the latter's momentum and is proportional to the square of its path--length, (\ref{dp dtau}) is linear in $p$.\\
As illustrated in previous works \cite{Dominguez:2008vd, Gubser:2006nz, Giecold:2009cg, Hatta:2008tx}, the mechanisms for momentum broadening appear to differ at weak and strong coupling, if $\mathcal{N} = 4$ provides any hint on QCD in the latter regime. We now explore how the momentum broadening coefficients are modified in an expanding plasma at strong coupling.

\subsection{Fluctuating trailing string and momentum diffusion}

This section is concerned with deriving the momentum broadening coefficients from fluctuations of the trailing string (\ref{X trail 2}). This was done in \cite{CasalderreySolana:2006rq, CasalderreySolana:2007qw, Giecold:2009cg, Gubser:2006nz} for the case of a static medium. For a review of jet quenching and momentum broadening in perturbative QCD and in AdS/CFT, the review \cite{CasalderreySolana:2007zz} is particularly recommended.\\
Writing
\beq\label{fluct ansatz}
X^1(t,z) = X^1_0(t,z) + \delta \xi^1(t,z) \ \ \ X^2(t,z) = \delta \xi^2(t,z) \ \ \ Y(t,z) = \delta y(t,z), 
\eeq
with fluctuating terms in the transverse and velocity directions, and inserting in the Nambu--Goto action after some algebra ultimately leads to the following expansion at quadratic order of the action:
\beq\label{NG pertb}
S_{NG} = -\frac{R^2/z_H}{2 \pi \alpha'} \int dt \ dz \frac{\sqrt{1 - \frac{3t_0}{2t} v^2}}{z^2} + \int dt \ dz P^{\alpha} \partial_{\alpha} \xi^1 \nonumber \\ - \frac{1}{2} \int dt \ dz T_{\delta y}^{\alpha \beta} \partial_{\alpha}\delta y \partial_{\beta}\delta y - \frac{1}{2} \int dt \ dz \sum_{i = 1,2} T_{\xi^{i}}^{\alpha \beta} \partial_{\alpha}\delta \xi^{i} \partial_{\beta}\delta \xi^{i},
\eeq
where
\beq\label{P alpha}
P^{\alpha} = -\frac{R^2/z_H^2}{2\pi \alpha'} \frac{\frac{3t_0}{2t}v}{\sqrt{1-\frac{3t_0}{2t} v^2}} \left(
\begin{array}{ccc}
z_H / (z^2 (1-z^4)) \\
1\\
\end{array}
\right), \eeq
\beq\label{T y}
T_{\delta y}^{\alpha \beta} = - \frac{R^2/z_H^2}{2\pi \alpha'} \frac{4t^2/9}{\sqrt{1-\frac{3t_0}{2t} v^2}}  \left(
\begin{array}{ccc}
\frac{z_H}{z^2} \frac{ \left[ 1-(1-\frac{3t_0}{2t}v^2)z^4 \right]}{(1-z^4)^2} & \frac{3t_0}{2t} \frac{v^2}{1-z^4} \\
\frac{3t_0}{2t} \frac{v^2}{1-z^4} & \frac{\left[ z^4 - (1-\frac{3t_0}{2t}v^2) \right]}{z_H z^2}  
\end{array}
\right), \eeq
and
\begin{align}\label{T xi}
T_{\xi^{2}}^{\alpha \beta} &= \left[ 1 - \frac{3t_0}{2t}v^2 \right] T_{\xi^{1}}^{\alpha \beta} \nonumber \\
&= - \frac{R^2/z_H^2}{2\pi \alpha'} \frac{3t_0/2t}{\sqrt{1-\frac{3t_0}{2t} v^2}}  \left(
\begin{array}{ccc}
\frac{z_H}{z^2} \frac{\left[ 1-(1-\frac{3t_0}{2t}v^2)z^4 \right]}{(1-z^4)^2} & \frac{3t_0}{2t} \frac{v^2}{1-z^4} \\
\frac{3t_0}{2t} \frac{v^2}{1-z^4} & \frac{\left[ z^4 - (1-\frac{3t_0}{2t}v^2) \right]}{z_H z^2}  
\end{array}
\right); \end{align}
   Making use of reparametrization--invariance on the world-sheet, these results translate into the following expressions :
\beq\label{NG pertb tau}
S_{NG} = -\frac{\sqrt{\lambda}}{2} \int d\tau \ dz \frac{\sqrt{1-\tilde{v}^2} T(\tau)}{z^2} + \int d\tau \ dz \tilde{P}^{\alpha} \partial_{\alpha} \xi^1 \nonumber \\ - \frac{1}{2} \int d\tau \ dz \tilde{T}_{\delta y}^{\alpha \beta} \partial_{\alpha}\delta y \partial_{\beta}\delta y - \frac{1}{2} \int d\tau \ dz \sum_{i = 1,2} \tilde{T}_{\xi^{i}}^{\alpha \beta} \partial_{\alpha}\delta \xi^{i} \partial_{\beta}\delta \xi^{i},
\eeq
with $\alpha, \beta$ running over $z, \tau$ and
\beq\label{P alpha tau}
\tilde{P}^{\alpha} = -\frac{\pi \sqrt{\lambda} T^2(\tau)}{2} \frac{\tilde{v}}{\sqrt{1-\tilde{v}^2}} \left(
\begin{array}{ccc}
\frac{1}{\pi T(\tau) (z^2 (1-z^4))} \\
1\\
\end{array}
\right), \eeq
\beq\label{T y tau}
\tilde{T}_{\delta y}^{\alpha \beta} = - \frac{R^2/z_H^2}{2\pi \alpha'} \frac{(\tau_0 \tau^2)^{\frac{2}{3}}}{\sqrt{1-\tilde{v}^2}}  \left(
\begin{array}{ccc}
\frac{1}{\pi T(\tau) z^2} \frac{\left[1-(1-\tilde{v}^2)z^4 \right]}{(1-z^4)^2} & \frac{\tilde{v}^2}{1-z^4} \\
\frac{\tilde{v}^2}{1-z^4} & \frac{\pi T(\tau) \left[z^4 - (1-\frac{3t_0}{2t}v^2) \right]}{z^2}  
\end{array}
\right), \eeq
and
\begin{align}\label{T xi tau}
\tilde{T}_{\xi^{2}}^{\alpha \beta} &= \left[1 - \tilde{v}^2 \right] \tilde{T}_{\xi^{1}}^{\alpha \beta} \nonumber \\ &= - \frac{\pi \sqrt{\lambda} T^2(\tau)}{2} \frac{1}{\sqrt{1-\tilde{v}^2}}  \left(
\begin{array}{ccc}
\frac{1}{\pi T(\tau) z^2}\frac{\left[1-(1-\tilde{v}^2)z^4 \right]}{(1-z^4)^2} & \frac{\tilde{v}^2}{1-z^4} \\
\frac{\tilde{v}^2}{1-z^4} & \frac{\pi T(\tau) \left[z^4 - (1-\tilde{v}^2) \right]}{z^2}  
\end{array}
\right). \end{align}
Recall that $\tau$ is related to $t$ through $\frac{t}{t_0} = \frac{3}{2} (\frac{\tau}{\tau_0})^{\frac{2}{3}}$. 
In the above, the temperature appears only through its local, proper--time dependent expression.
Let us now show that the momentum broadening coefficients are then formally the same as in \cite{CasalderreySolana:2007qw, Gubser:2006nz}.\\
Indeed, if one uses the second set $\tilde{T}^{\alpha \beta}$ of tensor densities, proper time derivatives of the temperature are discarded in the equations of motion, $\partial_{\alpha} T^{\alpha \beta} \partial_{\beta} \phi = 0$, given that they imply sub--leading $\mathcal{O}(\tau^{-4/3})$ contributions. Therefore, independent solutions to the equations of motion look the same as in \cite{Gubser:2006nz}\footnote{See also \cite{Giecold:2009cg} where they are labelled $\psi_{ret}(\omega, z)$ and $\psi_{adv}(\omega,z)$.}, with $T_0 \rightarrow T(\tau)$.\\
In $z, \tau$ coordinates the location $z_S$ of world--sheet horizon\footnote{A world--sheet horizon is generally determined from the zeroes of the polynomial factor appearing in front of the second AdS--radial derivative in the equations of motion. They determine the regular singular points of this equation. When $z = z_S$ the value of $\partial_z \phi$ at $z=z_S$ is determined from the equation of motion. This means that fluctuations of the string at $z < z_S$ are causally disconnected from those away from the location of the world-sheet horizon.} is $z_S = \sqrt[4]{1-\tilde{v}^2}$.\\
We are interested in the form of the Kruskal diagram in the $z$, $\tau$ coordinates with the JP asymptotic condition on the latter variable.\\
Keeping only the $z, \tau$ components, this reads $ds^2 = (R\pi T(\tau))^2/z^2 \left[-f d\tau^2 + dz^2 / \pi^2 T(\tau)^2 f \right]$, i.e. $ds^2 \simeq (R\pi T(\tau))^2/z^2 \left[-f d\tau^2 + [d(z / \pi T(\tau))\right]^2 / f]$.\\
At this order of the JP expansion the Kruskal coordinates are found as follows.\\
The null condition leads to $(\pi T(\tau))^2 (d\tau)^2 = \frac{(dz)^2}{f(z)^2}$, hence
\beq\label{Null cond}
\tau^{\frac{2}{3}} = \pm z_{*} + C.
\eeq
where $C$ labels a constant c--number and 
\beq\label{Z etoile}
z_{*} = \frac{1}{3 \pi T_0 \tau^{1/3}_0} [\arctan(z) + \frac{1}{2} \log(\frac{1+z}{1-z})].
\eeq
Introducing $\nu_{+} = \tau^{2/3} - z_{*}$ and $\nu_{-} = \tau^{2/3} + z_{*}$, the metric is written as
\beq\label{Metric nu plus minus}
ds^2 = - (\frac{3}{2} \pi R T_0 \tau_0^{1/3})^2 \frac{f(z)}{z^2} d\nu_{-} d\nu_{+}.
\eeq
$z$ and $\tau$ are given through the implicit equations
\beq\label{z tau implicit}
\left\{
\begin{array}{ll}
\tau = (\frac{\nu_{-} + \nu_{+}}{2})^{\frac{3}{2}},\\
\arctan(z) + \frac{1}{2} \log(\frac{1+z}{1-z}) = \frac{3}{2} \pi T_0 \tau_0^{1/3} (\nu_{-} - \nu{+}). 
\end{array} \right.
\eeq
It is then natural to introduce the variables $U$ and $V$, the Kruskal coordinate for this setting :
\beq\label{Kruska trail}
U = -e^{-3 \pi T_0 \tau_0^{1/3} \nu_{-}}, \ \ \ V = e^{3 \pi T_0 \tau_0^{1/3} \nu_{+}}. 
\eeq
$z$ and $\tau$ are then defined implicitly in those coordinates as
\beq\label{z tau Kruska}
\left\{
\begin{array}{ll}
-UV = \frac{1-z}{1+z} e^{-2 \arctan(z)},\\
-\frac{V}{U} = e^{6 \pi T(\tau) \tau} = e^{4 \pi T_0 t}. 
\end{array} \right.
\eeq
The Kruskal diagram is split into four quadrants by the curves $U = 0$ and $V = 0$. $UV = 0$ still yields $z = 1$ and $\tau$ is given by $\log(-\frac{V}{U}) = 6\pi T(\tau) \tau$, so that $V = 0$, resp. $U = 0$, still corresponds to $\tau = -\infty$ or $t = -\infty$, resp. $+\infty$. This conclusion is supported by \cite{Figueras:2009iu}, Fig. 1, where they show that the BF geometry is a regular black hole spacetime. The apparent and event horizons were found at various orders in the JP metric and they tend to a common slowly varying line in the $z$--$\tau$ plane when $\tau >> 1$.\\
The trailing string solution (\ref{X trail 2}) is cast in the form
\beq\label{X trail Kruska}
X^1_0(t,z) = x^1_0 + \frac{v}{2 \pi T_0} \log(V) + \frac{v}{\pi T_0} \arctan(z),
\eeq
which explicitly shows that the trailing string is regular at the horizon between the upper and the right quadrants. A state of the system is prepared at $Re \ t = -\infty$, which corresponds to the singularity at $V = 0$, is propagated along $Im \ t = 0$, and back along $Im \ t = - \sigma$, for some constant $\sigma$ in the Schwinger--Keldysh path.\\
This suggests that all the analysis exposed in ~\cite{Giecold:2009cg, Gubser:2006nz} is directly applicable to the current problem, with the proviso that the world--sheet horizon in the ${t-z}$ coordinates is now time--dependent, $z_S = \sqrt[4]{1 - \frac{3t_0}{2t}v^2}$. Indeed the equations of motion obtained from (\ref{P alpha}), (\ref{T y}) and (\ref{T xi}) are the same as those found in the above references, given that the additional time factors should be neglected at the order of the JP expansion one is dealing with. The additional time--dependence of the location to the world--sheet horizon is accounted for by noticing that it turns out to be subleading compared to the decorrelation time.\\
All in all, after going back to the Bjorken frame as was done at the end of Section 2, this yields
\beq\label{qqq hat trail}
\hat{q}_{\xi^2}(\tau) = \pi \sqrt{\lambda} T^3(\tau) \frac{ \sqrt{ \gamma( \tilde{v} ) } }{\tilde{v}} \ \ \ \hat{q}_{\xi^1}(\tau) = \pi \sqrt{\lambda} T^3(\tau) \frac{\gamma(\tilde{v})^{5/2}}{\tilde{v}} \ \ \ \gamma(\tilde{v}) = 1/\sqrt{1-\tilde{v}^2}.
\eeq
This corresponds to a stochastic force in a Langevin equation satisfying
\beq\label{Lang stoch trail}
\frac{d\tilde{p}_i}{d\tau} = F_i \ \ \ \langle F_i(\tau_1) F_j(\tau_2) \rangle = \delta_{ij} K_i(\tau_1, \tau_2), \ \ i,j = 1,2
\eeq 
From $\hat{q} = \langle p^2 \rangle / l$ --- $l$, the path length travelled by the quark in the plasma proper frame, being large enough that the memory short range correlations is not taken into account but large enough that the quark has not departed significantly from its initial trajectory --- this gives 
\beq\label{q hat Gubs}
\hat{q}_i = \frac{1}{\tilde{v}} \int d\tau K_i(\tau, 0),
\eeq
as defined in, e.g. \cite{Gubser:2006nz}.\\
Note however that (\ref{qqq hat trail}) bears no relation to jet--quenching. The longitudinal and transverse parts do not match at finite momentum. The momentum broadening coefficients have been written in this varnished form to suggest a similarity to the local jet--quenching parameter appearing in perturbative QCD calculations with a high energy quark moving in an expanding plasma.

\subsection*{Acknowledgments}
I would like to thank C. Gombeaud for discussion and particularly Doctor E. Iancu and Professor A. H. Mueller for comments on the manuscript. This work was supported in part by Contrat de Formation par la Recherche, CEA--Saclay.

\end{document}